# Impact of Dielectric Environment on Trion Emission from Single-Walled Carbon Nanotube Networks

Sonja Wieland, Abdurrahman Ali El Yumin, Jan M. Gotthardt, and Jana Zaumseil*

**ABSTRACT:** Trions are charged excitons that form upon optical or electrical excitation of low-dimensional semiconductors in the presence of charge carriers (holes or electrons). Trion emission from semiconducting single-walled carbon nanotubes (SWCNTs) occurs in the near-infrared and at lower energies compared to the respective exciton. It can be used as an indicator for the presence of excess charge carriers in SWCNT samples and devices. Both excitons and trions are highly sensitive to the surrounding dielectric medium of the nanotubes, having an impact on their application in optoelectronic devices. Here, the influence of different dielectric materials on exciton and trion emission from electrostatically doped networks of polymer-sorted (6,5) SWCNTs in top-gate field-effect transistors is investigated. The observed differences of trion and exciton emission energies and intensities for hole and electron accumulation cannot be explained with the polarizability or screening characteristics of the different dielectric materials, but they show a clear dependence on the charge trapping properties of the dielectrics. Charge localization (trapping of holes or electrons by the dielectric) reduces exciton quenching, emission blue-shift and trion formation. Based on the observed carrier type and dielectric material dependent variations, the ratio of trion to exciton emission and the exciton blue-shift are not suitable as quantitative metrics for doping levels of carbon nanotubes.

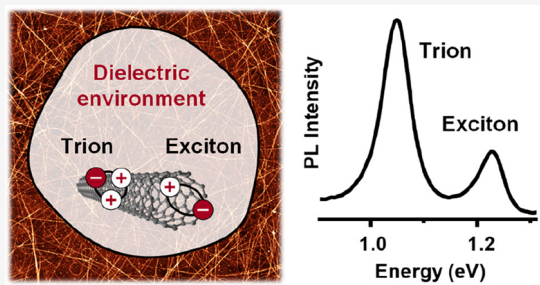

## ■ INTRODUCTION

The unique optical properties of semiconducting single-walled carbon nanotubes (SWCNTs) are governed by their quasi one-dimensional geometry and the resulting strongly bound excitons.[1−3] The main excitonic $E_{11}$ absorption and emission of individual SWCNTs is very narrow (<20 meV)[4,5] and occurs at diameter-dependent energies in the near-infrared (NIR).[6,7] Substantial progress in the purification and sorting of SWCNTs[8,9] has enabled the observation and investigation of a second species of excited-state quasiparticles in doped carbon nanotubes, i.e., trions (see Figure 1a). Trions ($T^+/T^-$) can be considered as positively or negatively charged excitons that are formed upon optical or electrical excitation in the presence of chemically,[10,11] electrochemically,[12−15] or electrostatically[16−18] induced excess charge carriers. Trion emission from SWCNTs is red-shifted by 100−200 meV compared to the corresponding $E_{11}$ transition (see Figure 1b). This energetic separation ($\Delta E_{E_{11}-T}$) is the sum of the singlet−triplet exciton exchange splitting and the trion binding energy that scale inversely with the squared nanotube diameter and the diameter, respectively.[10,12] While excess charge carriers promote the creation of trions (charged excitons), they also cause nonradiative decay of excitons due to Auger-type quenching.[19,20] Consequently, the observation of trion emission and its intensity relative to excitonic emission ($T/E_{11}$) should be indicative of SWCNT doping levels and might be used as an optical probe for carrier density.[21] Further, trions are often observed but highly undesired in electroluminescent devices,[18,22,23] as they lead to additional emission peaks and thus reduced spectral purity as well as lower external quantum efficiency. Hence, their suppression or at least control is important for carbon nanotube-based optoelectronics.

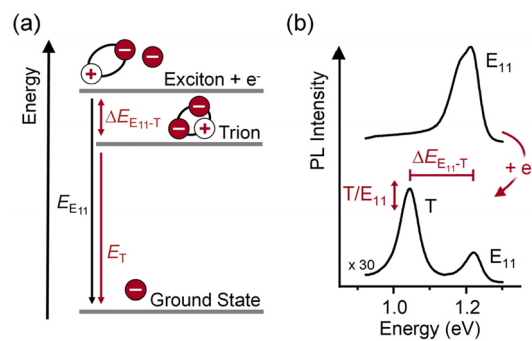

**Figure 1.** (a) Schematic energy diagram of excitons and trions in doped semiconducting SWCNTs (here: negative trions by electron doping). (b) Schematic PL spectra of a (6,5) SWCNT network in the neutral (top) and electrostatically doped state (bottom). The $E_{11}$ emission is reduced, and a trion peak emerges for the doped state.

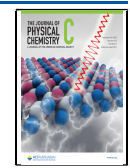

Due to their quasi one-dimensional structure and thus limited screening, the optical and electronic properties of SWCNTs, including exciton and trion energies, are highly sensitive to their dielectric environment. Interactions with their surroundings are evident in the broadening of the $E_{11}$ transition upon deposition of SWCNTs onto different substrates, which also adds substantial inhomogeneity.[4,24,25] Theoretical and experimental studies on the impact of the dielectric environment of individual SWCNTs[26,27] and nanotube dispersions[28−30] typically found a decrease of photoluminescence (PL) quantum yield and lower transition energies with increasing dielectric constants due to screening effects. Further, defects and charged impurities on or in the direct vicinity of SWCNTs have been found to be major sources of exciton quenching.[31,32] They also promote trion formation.[11,19,33] Trion emission has also been investigated experimentally in a variety of environments (e.g., in ionic liquid or in aqueous dispersion)[33−36] and for different SWCNT chiralities;[12] however, systematic studies on their precise interactions with their dielectric environment are still lacking. Theory predicts a reduction of the trion binding energy with an increasing dielectric constant of the nanotube environment.[37] This notion was indeed corroborated for trions in free-standing, individual SWCNTs that were hole-doped either electrostatically or chemically.[27,38]

The dielectric environment is not only important for the optical properties of SWCNTs but also directly affects their charge transport. Early theoretical studies suggested that surface polar phonon scattering substantially reduces the carrier mobility in individual carbon nanotubes and, thus, high dielectric constant materials such as hafnia or zirconia should be detrimental for achieving high carrier mobilities.[39] In analogy to organic semiconductors,[40] the impact of dipolar disorder created by very polar, high dielectric constant materials on the carrier mobilities in random networks of SWCNTs was studied.[41] However, the differences between hole and electron transport depending on dielectric could not be resolved conclusively, and additional effects such as trap states created during the deposition process or present in the dielectric make analysis of such devices more complex.[42]

An accessible and commonly used nanotube species for both spectroscopic[43−45] as well as charge transport studies[46−49] are (6,5) SWCNTs, which can be obtained in sufficient quantities through various separation techniques including polymer wrapping using poly[(9,9-dioctylfluorenyl-2,7-diyl)-alt-(6,6′-(2,2′-bipyridine))] (PFO-BPy).[50] The relatively large band gap and narrow $E_{11}$ excitonic transition of (6,5) nanotubes around 1.24 eV enable easy detection even of further red-shifted trion emission. Holes and electrons can be injected and accumulated in (6,5) SWCNT networks with nearly equal efficiency making them a suitable material for the investigation of both positive and negative trions.[13] Chemical, electrochemical, and electrostatic doping of (6,5) SWCNT networks has been demonstrated. However, electrostatic doping in a field-effect transistor (FET) structure has the clear advantage of tuning the carrier density without moving or introducing any counterions or chemical dopants that may alter the energy landscape of excitons and trions in addition to a given dielectric environment. The dielectric environment of nanotubes in an FET can be controlled and varied over a wide range through the gate dielectric material that is in direct contact with the active channel. Moreover, it is possible to continuously adjust the charge carrier density and hence the probability of trion formation by the applied gate voltage ($V_g$) while also monitoring the number of mobile carriers via the current flow (drain current, $I_d$) between source and drain electrode for a constant drain voltage ($V_d$).[51]

Here, we systematically investigate the influence of the dielectric environment on exciton and trion emission from electrostatically doped networks of semiconducting (6,5) SWCNTs in top-gate FETs by employing five different polymeric and inorganic gate dielectrics with static dielectric constants $\varepsilon$ ranging from 1.9 to over 20 and with different charge trapping properties. The role of charge traps for trion emission is further tested by treating the (6,5) SWCNT networks with 1,2,4,5-tetrakis(tetramethylguanidino)benzene (ttmgb) to efficiently remove electron traps.

## ■ METHODS

**Preparation of (6,5) SWCNT Dispersions.** Nearly monochiral (6,5) SWCNT dispersions were obtained by shear-force mixing (Silverson L2/Air, 10230 rpm) of 50 mg of CoMoCat raw material (Sigma-Aldrich, MKCJ7287) in a solution of 65 mg of PFO-BPy (American Dye Source, Inc., $M_w$ = 40 kg mol$^{-1}$) in 140 mL of anhydrous toluene for 72 h at 20 °C as previously described.[50] The PFO-BPy-wrapped (6,5) SWCNTs were separated from nonexfoliated material by two consecutive centrifugation steps (45 min at 60 000g, Beckman Coulter Avanti J26XP centrifuge). After the second centrifugation step, the supernatant was filtered through a syringe filter (Whatman PTFE membrane, pore size 5 μm) to yield a polymer-rich stock dispersion of (6,5) SWCNTs. The purity of the (6,5) SWCNTs was confirmed by absorption (Cary 6000i UV−vis−NIR spectrometer, Varian Inc.) and Raman spectroscopy (InVia Reflex, Renishaw plc), see Figure S1, Supporting Information. For device fabrication, the stock dispersion was vacuum-filtered through a polytetrafluoroethylene membrane (Merck Millipore, JVWP, pore size 0.1 μm), and the filter cake was submerged three times in 10 mL of toluene for 10 min at 80 °C to remove excess polymer. The (6,5) SWCNTs were redispersed by ultrasonication for 30 min in toluene. The toluene volume was adjusted to yield an absorbance of 5 at the $E_{11}$ transition for 1 cm path length.

**Device Fabrication.** Top-gate/bottom-contact SWCNT network transistors were fabricated on cleaned glass slides (sodium-free aluminum borosilicate glass, Schott AF32 eco, 20 × 25 mm$^2$, thickness 0.3 mm). The interdigitated bottom electrodes (channel length $L$ = 10 and 20 μm with channel width $W$ = 10 mm or $L$ = 40 μm with $W$ = 5 mm) were patterned by photolithography using a double-layer photoresist (LOR5B (MicroChem)/ MICROPOSIT S1813 (Dow Chemical)). An adhesion layer of 2 nm of chromium and 30 nm of gold was deposited by electron beam evaporation, followed by lift-off in N-methyl-2-pyrrolidone (NMP, Sigma-Aldrich, 99%) and a cleaning step with ultrasonication in acetone and 2-propanol. Prior to deposition of (6,5) SWCNTs, the substrates were treated in a UV/ozone cleaner for 10 min, rinsed with 2-propanol, blow-dried with nitrogen, and annealed for 5 min at 100 °C. To obtain homogeneous networks, the (6,5) SWCNTs were spin-coated onto the substrates three times

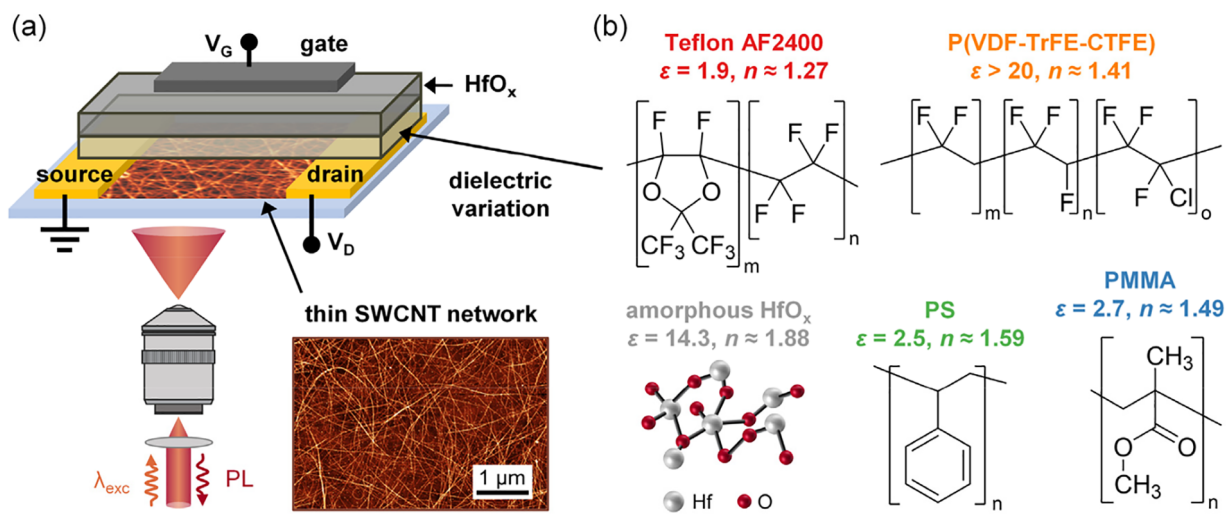

**Figure 2.** (a) Top-gate transistor layout and PL measurement setup with atomic force micrograph of a spin-coated (6,5) SWCNT network. (b) Molecular structures of the dielectric materials in contact with the SWCNT network including their static dielectric constants $\varepsilon$ and refractive indices $n$.

(30 s, 2000 rpm). After each spin-coating step, the substrates were heated for 2 min at 90 °C. To remove excess polymer, the samples were washed with tetrahydrofuran (VWR International, analytical grade) and 2-propanol and subsequently heated for 4 min at 90 °C. The linear density of the resulting SWCNT networks was 30 nanotubes/μm with an average SWCNT length of 1.2 ± 0.4 μm, as determined by atomic force microscopy (AFM, Bruker Dimension Icon). SWCNTs outside the channel area were removed by oxygen plasma etching while SWCNTs within the channel were protected with photolithographically patterned photoresist as described above. After the subsequent lift-off in NMP and annealing for 45 min at 300 °C in dry nitrogen, selected samples were doped by submerging the substrates in a 3 or 6 g·L$^{-1}$ solution of 1,2,4,5-tetrakis(tetramethylguanidino)benzene (ttmgb)[52] in anhydrous toluene for 20 min, followed by annealing for 20 min at 150 °C. Polymeric dielectric films with thicknesses of about 50 nm were deposited in a dry nitrogen atmosphere by spin-coating and successive annealing: Teflon AF2400 (poly-[4,5-difluoro-2,2-bis(trifluoromethyl)-1,3-dioxole-co-tetra-fluoroethylene], Sigma-Aldrich, 53 nm from 20 g·L$^{-1}$ solution in FluorInert FC-40 (3M)); P(VDF-TrFE-CTFE) (poly-(vinylidene fluoride-co-trifluoroethylene-co-chlorotrifluoroethylene) Piezotech RT TS, Piezotech Arkema, 52 nm from 10 g·L$^{-1}$ in n-butanone); PMMA (poly(methyl methacrylate), PolymerSource, $M_W$ = 315 kg·mol$^{-1}$, syndiotactic, 50 nm from 20 g·L$^{-1}$ in $n$-butyl acetate); PS (polystyrene, $M_w$ = 230.4 kg·mol$^{-1}$, atactic, 50 nm from 20 g·L$^{-1}$ in $n$-butyl acetate). 60 nm of HfO$_x$ were deposited via atomic layer deposition (ALD, Ultratech Savannah S100) at 100 °C with tetrakis-(dimethylamino)hafnium (Strem Chemicals Inc.) and water as precursors either as the top layer of a hybrid dielectric or as a single dielectric layer directly on the SWCNTs. Thermal evaporation of 30 nm thick silver gate electrodes through a shadow mask completed the devices. Due to the conformal encapsulation with ALD-HfO$_x$, all devices were air-stable.

**Electrical and Characterization.** Current–voltage characteristics were recorded in a dry nitrogen atmosphere with a semiconductor parameter analyzer (Agilent 4155C). Areal capacitances ($C_i$) were determined using an LCR meter (Agilent 4980A) at 1 kHz. Linear hole and electron mobilities were calculated from the forward gate voltage sweeps (from off to on at $V_d = -0.1$ V) of FETs with the longest channel lengths $L = 40$ μm ($W = 5$ mm) to minimize the impact of contact resistance using the equation $\mu_{\text{lin}} = \frac{\partial I_d}{\partial V_g} \cdot \frac{L}{W \cdot C_i \cdot V_d}$. Areal charge carrier densities $Q = C_i \cdot V_{\text{eff}}$ were calculated with the effective voltage $V_{\text{eff}} = V_g - V_{\text{max. PL}}$ to account for dielectric-specific shifts in the off-voltage (i.e., the voltage of maximum PL intensity $V_{\text{max. PL}}$). Trap densities were estimated from the subthreshold swing $S$ according to $N = \left[\frac{S \cdot e}{kT \ln 10} - 1\right] \cdot \frac{C_i}{e^2}$ with the elementary charge $e$, the temperature $T$, and the Boltzmann constant $k$.[53]

**Optical Characterization.** Photoluminescence (PL) images and spectra were acquired from FETs (dielectric variation devices: $L = 20$ μm, ttmgb devices: $L = 10$ μm) through the glass substrate with a near-infrared ×20 objective (Olympus, NA 0.4), a Princeton Instruments IsoPlane SCT 320 spectrometer, and a thermo-electrically cooled InGaAs camera (NIRvana 640ST, Princeton Instruments, 512 × 640 pixels). The excitation laser beam (640 nm, continuous-wave, OBIS, Coherent) was expanded with a plano convex lens (focal length $f = 125$ mm) in front of the objective, and PL spectra were recorded and averaged over an area of 700 μm$^2$. Scattered laser light was blocked using a 700 and an 850 nm long-pass filter. The PL spectra were corrected for the detector sensitivity and the absorption of the optics in the detection path. Spectra were fitted with three Lorentzians after Jacobian conversion[54] of wavelength to energy scale. PL quenching was induced by applying a constant gate voltage using a Keysight B1500A semiconductor parameter analyzer. A small drain voltage of $V_d = 0.01$ V was applied during PL quenching experiments to record current flow without significantly altering the charge carrier distribution along the channel.

## RESULTS AND DISCUSSION

To understand the impact of the dielectric environment on the emission properties of electrostatically doped carbon nanotubes, spin-coated (6,5) SWCNT networks were integrated in bottom-contact/top-gate FETs as shown in Figure 2a. The network density was chosen to be relatively sparse (∼30 nanotubes/μm) to guarantee that all SWCNTs were in direct contact with the solid gate dielectric on top but also dense enough to yield sufficient percolation for uniform charge transport.[55] Some inhomogeneities of the network density were expected, e.g., due to bundling of SWCNTs during the deposition process, and compensated for as discussed below. The nanotubes were also in contact with the glass substrate (sodium-free aluminum borosilicate glass, static dielectric constant 5.1), which formed another component of the dielectric environment but remained constant throughout all experiments.

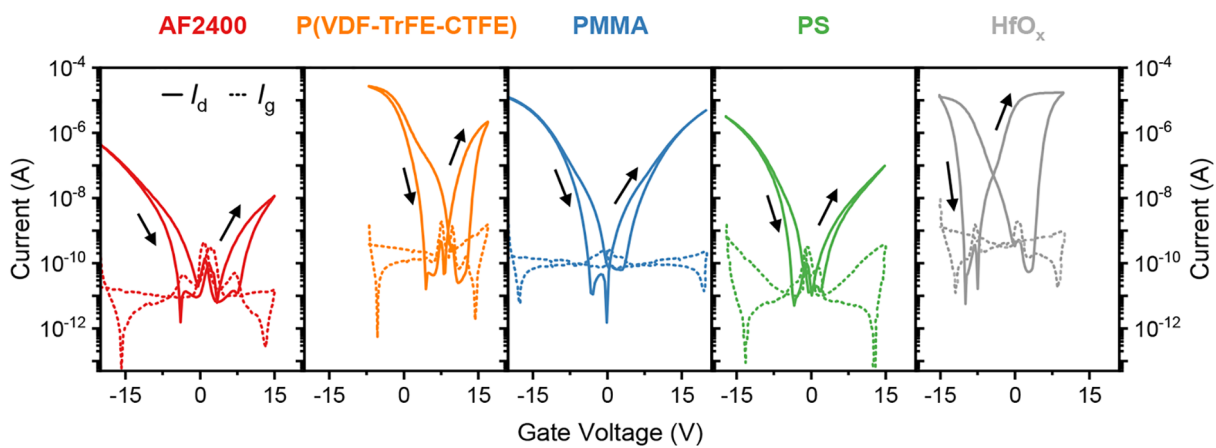

**Figure 3.** Ambipolar transfer characteristics of FETs with different dielectrics in contact with random (6,5) SWCNT networks (channel length $L$ = 20 μm, channel width $W$ = 10 mm) acquired at $V_d$ = 0.01 V.

**Table 1. Linear Hole ($\mu_{lin,h}$) and Electron ($\mu_{lin,e}$) Mobilities Extracted from FETs with Different Gate Dielectrics and Their Corresponding Static Dielectric Constants $\varepsilon$**[a]

|  | AF2400 | P(VDF-TrFE-CTFE) | PMMA | PS | HfO$_x$ |
|---|---|---|---|---|---|
| static dielectric constant $\varepsilon$ | 1.9 | >20 | 2.7 | 2.5 | 14.3 |
| $\mu_{lin,h}$ (cm$^2$ V$^{-1}$ s$^{-1}$) | 8.0 ± 0.8 | 13.7 ± 1.2 | 15.0 ± 1.2 | 7.1 ± 0.6 | 3.8 ± 0.3 |
| $\mu_{lin,e}$ (cm$^2$ V$^{-1}$ s$^{-1}$) | 2.0 ± 0.1 (*) | 1.6 ± 0.3 | 13.5 ± 0.8 | 2.7 ± 0.2 | 5.2 ± 0.4 |

[a]The asterisk (*) indicates that the maximum mobility was not reached before device breakdown.

A range of materials with different static dielectric constants $\varepsilon$, refractive indices $n$, and molecular structures was employed as the gate dielectric in direct contact with the SWCNTs (see Figure 2b). The thickness (50 nm) of the homogeneous, pinhole-free layers (see Supporting Information, Figure S2) of the polymeric dielectrics Teflon AF2400, poly(vinylidene fluoride-co-trifluoroethylene-co-chlorotrifluoroethylene) (P(VDF-TrFE-CTFE)), poly(methyl methacrylate) (PMMA), and polystyrene (PS) was chosen to exclude any direct impact of the hafnia (HfO$_x$) layer that was added on top by atomic layer deposition (ALD) for encapsulation. In addition to these hybrid polymer/oxide dielectrics, FETs with a single layer of amorphous ALD-HfO$_x$ as the gate dielectric were fabricated. All dielectric materials were chosen based on their successful previous application in either SWCNT[41,56] or organic semiconductor FETs,[57,58] their dielectric constant, their different molecular structures (e.g., fluorinated, with polar or aromatic side groups, etc., see Figure 2b), and their processability from solution for the polymer layers. All devices were prepared at the same time under identical conditions with the same batch of polymer-sorted (6,5) SWCNTs. Hence, any differences in charge transport and emission characteristics should result from the specific interactions of the different gate dielectric materials with the nanotubes. We note that, despite extensive rinsing of the nanotube networks, some wrapping polymer remains on the nanotubes (covering less than 20% of the surface)[59] and thus becomes part of the dielectric environment. However, this effect should also be constant within one batch of polymer-sorted nanotubes.

Figure 3 shows representative transfer characteristics of (6,5) nanotube FETs with different gate dielectric materials at low drain voltages. Note that the applied gate voltage ranges had to be varied and adjusted to reach the respective mobility maxima (see below) depending on the different dielectric constants of the materials and hence different areal capacitances (see Table S1, Supporting Information). All measured FETs exhibited ambipolar transport with some differences between the linear field-effect mobilities ($\mu_{lin}$) of holes and electrons as presented in Table 1. The carrier mobilities were gate voltage-dependent due to the one-dimensional density of states of SWCNTs as shown previously.[46] Gate leakage currents ($I_g$) were consistently low and did not affect the measurements. The maximum mobility values were extracted without correcting for contact resistance, which usually leads to a mobility underestimation of 20−30%. Different degrees of non-Ohmic charge injection were observed in the FET output characteristics (see Supporting Information, Figure S3), indicating that the dielectric materials also influence the charge injection barriers to some degree.[57] A detailed study of carrier mobilities depending on the employed dielectrics would require gated four-point probe measurements;[46] however, here we will only discuss the general trends and focus on the spectroscopic properties of the doped nanotubes.

FETs with the fluorinated polymers AF2400 and P(VDF-TrFE-CTFE) and those with PS consistently exhibited lower electron than hole mobilities (see Table 1). Fluorinated polymers are known for their electron trapping properties (e.g., as electrets),[60,61] which might explain the lower electron mobilities. Additional indicators of electron trapping in P(VDF-TrFE-CTFE) devices were the shift of the onset voltages toward positive gate voltages as well as a large drain current hysteresis, which did not vanish upon sweeping only toward positive gate voltages, i.e., electron accumulation (see Supporting Information, Figure S4). FETs with PS did not exhibit such a pronounced onset voltage shift, and current hysteresis vanished for limited gate voltage sweeps (only hole accumulation or only electron accumulation), indicating no significant charge trapping for either carrier (i.e., similar to devices with PMMA) despite the differences in mobility.

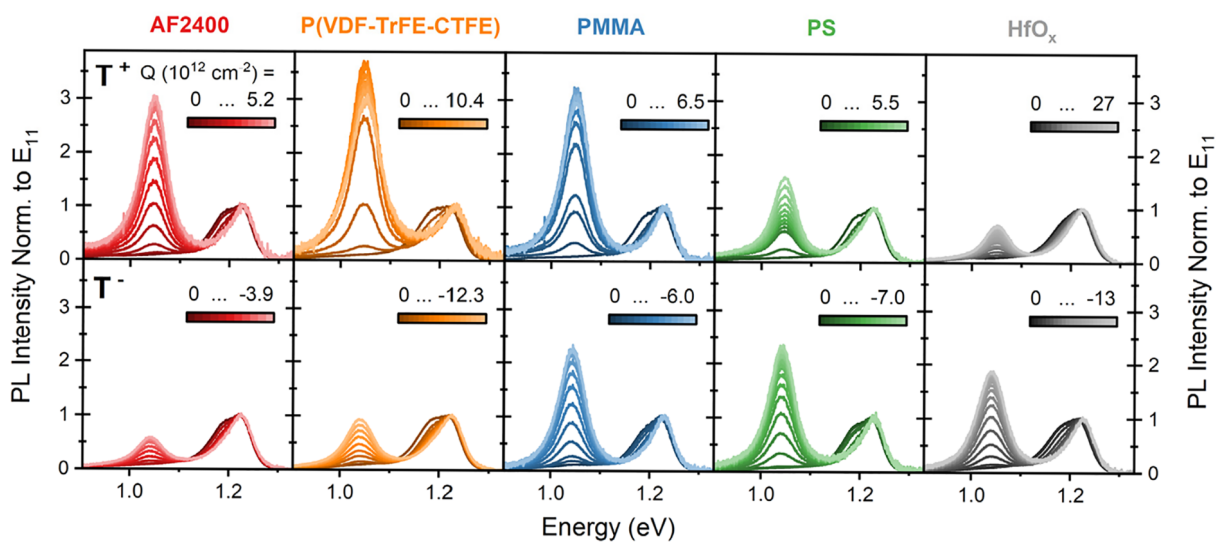

**Figure 4.** Charge carrier density-dependent evolution of PL spectra normalized to $E_{11}$ for hole (positive trions $T^+$, top row) and electron (negative trions $T^-$, bottom row) accumulation in FETs with different gate dielectrics in contact with the SWCNT network.

FETs with a single-layer $HfO_x$ dielectric and with the PMMA hybrid dielectric showed fairly balanced ambipolar transport. For devices with PMMA, the drain currents were nearly symmetrical around $V_g$ = 0 V, hole and electron mobilities were similar, and hysteresis was low. These near-ideal properties of hybrid dielectric layers of PMMA/$HfO_x$ have been widely employed for ambipolar FETs with different SWCNT networks.[42,46,55] In contrast to that, the transfer curves of FETs with only $HfO_x$ were strongly shifted toward negative gate voltages. Further, a significant hysteresis was observed that was also present in the hole-only, negative gate voltage sweeps (see Supporting Information, Figure S4). These observations can be explained with the known hole-trapping properties of $HfO_x$ that have been exploited previously to create pure n-channel SWCNT transistors.[62]

Overall, the transfer characteristics of these different FETs already indicate a strong impact of the various dielectrics on hole and electron transport. However, no clear trend of the extracted carrier mobilities as a function of the dielectric constant of the gate insulator and the induced dipolar disorder could be derived.

Despite the discussed differences in hole and electron mobilities, it was possible to reproducibly accumulate holes as well as electrons with gate voltage-tunable densities in all FETs with (6,5) SWCNT networks and study their gate voltage-dependent photoluminescence (PL) properties. Importantly, these measurements enabled the investigation of both positive and negative trions within one sample and one dielectric material system. Note however that the highest achievable carrier densities varied for each sample due to different dielectric breakdown limits.

To study exciton and trion emission, photoluminescence spectra were recorded under different applied gate voltages. The excitation laser beam was expanded, and PL spectra were averaged over an area of 700 $\mu m^2$ to remove any spot-to-spot variation and account for the inevitable SWCNT network density inhomogeneities that have an impact on the observable trion to exciton ratios (see Supporting Information, Figure S5). Note that both the oscillator strength for absorption (including higher transitions such as $E_{22}$) and emission efficiency of formed $E_{11}$ excitons are reduced with increasing charge carrier concentration.[11,17] The maximum PL intensity of any SWCNT network should thus correspond to the state with the lowest overall carrier density ($Q$), here defined as the zero-charge state, $Q$ = 0. In agreement with trapping-induced onset voltage shifts as observed in the transfer characteristics (see Figure 3), this state was reached at positive gate voltages for the fluorinated polymer dielectrics (AF2400 and P(VDF-TrFE-CTFE)) and at negative gate voltages for hafnia and PS (see Supporting Information, Table S2). The calculated charge carrier densities induced in the channel by the applied gate voltage (as the sum of static and mobile charges) were corrected for these gate voltage shifts (see Methods).

In the uncharged (neutral) state, the PL spectra of all devices exhibited a broad excitonic peak between 1.220 and 1.226 eV (see Figure 4). This broad emission peak width of nanotube networks integrated in FETs compared to dispersions results from device processing, which includes multiple heating steps (see Methods). Defect-related emission such as the Y-band shoulder[63] at energies between 1.180 and 1.189 eV was prominent and considered by fitting all spectra with two Lorentzian peaks to extract the $E_{11}$ exciton energy (see Methods and Supporting Information, Figure S6). The Y-band feature became less pronounced upon doping as evident from the normalized spectra in Figure 4, indicating that this defect-related emission was quenched at lower doping levels than the $E_{11}$ emission as also found for $sp^3$-defect emission in other studies.[64] Since the Y-defect states are nevertheless of excitonic nature and separate treatment of $E_{11}$ and Y-band emission did not alter the observed trends, the combined emission intensities are considered in the following.

The fitted $E_{11}$ peak energies in the neutral state decreased from AF2400, PS, PMMA, and P(VDF-TrFE-CTFE) to $HfO_x$ (see Supporting Information, Figure S7a), roughly following the expected trend based on increased dielectric screening.[26−29,65] Likewise, the emission peaks were broadest for P(VDF-TrFE-CTFE) and $HfO_x$, reflecting the highest dielectric constants and thus most pronounced dipolar disorder (see Supporting Information, Figure S7b).

To assess the evolution of exciton and trion emission depending on charge carrier density, i.e., applied gate voltage, all PL spectra recorded from the channel area were normalized to the $E_{11}$ peak (see Figure 4). For all FETs and dielectrics, the excitonic emission decreased nonlinearly (see Figure 5a and

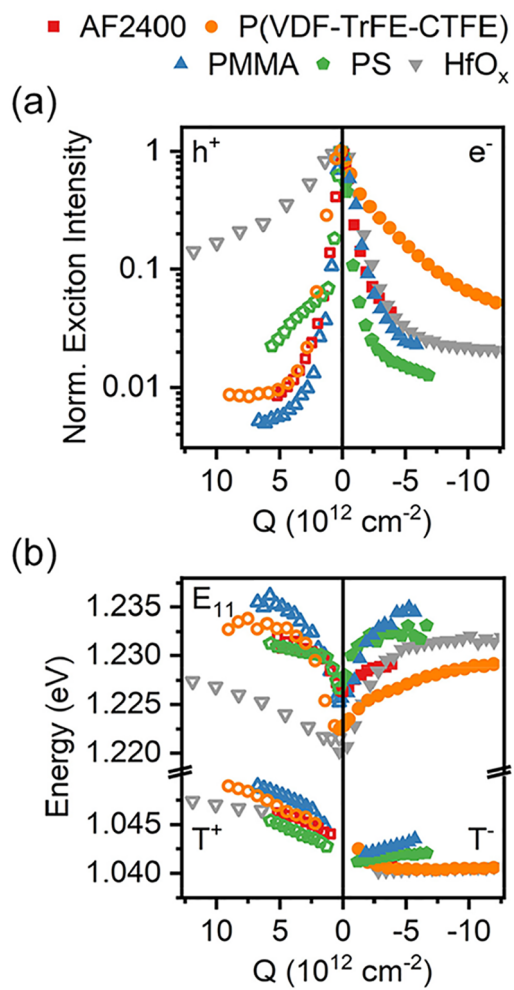

**Figure 5.** Evolution of (a) exciton quenching and (b) excitonic $E_{11}$ and trion ($T^+/T^-$) emission energies with charge carrier density Q for hole (open symbols) and electron (solid symbols) accumulation.

Supporting Information, Figure S8 for complete Q range) upon charge accumulation mainly due to Auger-type quenching.[19,20] Interestingly, the degree of quenching was different for positive and negative charges even within the same device; hence, neither the total number of charges nor the dielectric constant of the gate dielectric can be the only determining factor for the quenching efficiency. The most pronounced charge carrier polarity difference was observed for devices with a P(VDF-TrFE-CTFE) dielectric, which showed much lower exciton quenching for electrons than for holes. In contrast to that, the exciton quenching was weaker upon hole accumulation compared to electron accumulation in FETs with $HfO_x$. Since these two device types also showed the greatest hysteresis induced by electron and hole trapping, respectively, we may assume that trapped charge carriers contribute less to exciton quenching than mobile charge carriers. This notion is in agreement with the symmetrical quenching efficiency of vacuum-gated free-standing SWCNTs demonstrated by Yoshida et al.[16]

Carrier-induced exciton quenching in nanotubes is always accompanied by a blue-shift of the $E_{11}$ peak (see Figure 4 and Figure 5b, see Supporting Information, Figure S8 for complete Q range) as a result of a reduced exciton binding energy caused by electrostatic doping.[15,66] Similar to the exciton quenching efficiency, the magnitude of the $E_{11}$ peak shift depends not only on the dielectric material in contact with the SWCNTs but also on the polarity of the induced charge carriers. This observation indicates again that additional factors must play a role. The maximum $E_{11}$ energy shift for a specific dielectric was larger when carrier trapping of the dielectric was less pronounced, e.g., for hole accumulation in an FET with the electron-trapping P(VDF-TrFE-CTFE) dielectric (see Supporting Information, Table S3).

As discussed above, trion emission emerges in the PL spectra as exciton emission is quenched. The observed trion peak energies are about 1.045 and 1.040 eV for hole and electron accumulation, respectively. They are comparable to the values reported by Jakubka et al. for iongel-gated (6,5) SWCNT networks.[13] Trions also exhibit a doping-dependent blue-shift (see Figure 5b, see Supporting Information, Figure S8 for complete Q range), which probably arises from a reduction of the trion binding energy. This shift is smaller than the $E_{11}$ peak shift, which is in agreement with findings by Yoshida et al., who argued that an increase in singlet−triplet exciton splitting upon doping might be the underlying cause.[16]

A closer look at the evolution of the normalized spectra in Figure 4 and the trion to exciton intensity ratios versus carrier densities in Figure 6a (see Supporting Information, Figure S8 for complete Q range) reveals that, for many FETs, a maximum is reached before the trion to exciton ratio decreases again at even higher carrier concentrations. This maximum might be explained by the reduced $E_{22}$ absorption cross section and further increasing Auger quenching at higher carrier densities, which results in a lower number of excitons that are available for trion formation.[13,16] As shown in Figure 6a, the maximum trion/exciton ratio was not reached for all devices due to limited maximum induced carrier concentrations before dielectric breakdown occurred. Positive trions in FETs with PS gate dielectric were excluded from the following analysis, as they appeared to be still far from reaching the maximum value.

Despite these shortcomings, some clear trends of the maximum values of trion/exciton ratios for different gate dielectrics and for hole versus electron accumulation could be observed. The maximum trion/exciton ratios depended on the dielectric material as well as carrier polarity and occurred at different carrier densities (see Figure 6a). While maxima were generally reached for hole and electron densities below $10^{13}$ $cm^{−2}$, the maximum positive trion/exciton ratio for FETs with $HfO_x$ occurred at an exceptionally high hole concentration of $2.5 \times 10^{13}$ $cm^{−2}$ (see Supporting Information, Figure 8d). No clear trend with dielectric constant or refractive index of the gate dielectric could be found, but similar to the $E_{11}$ shift and exciton quenching, a correlation of the maximum trion/exciton ratio with the charge trapping properties of the dielectrics was implied. Electron-trapping fluorinated dielectrics yielded the lowest negative trion/exciton ratios, while the hole-trapping HfO$_x$ yielded the lowest positive trion/exciton ratio. For the commonly used dielectric PMMA, which also showed the least charge trapping and most balanced carrier mobilities, the positive and negative trion/exciton ratios were fairly similar.

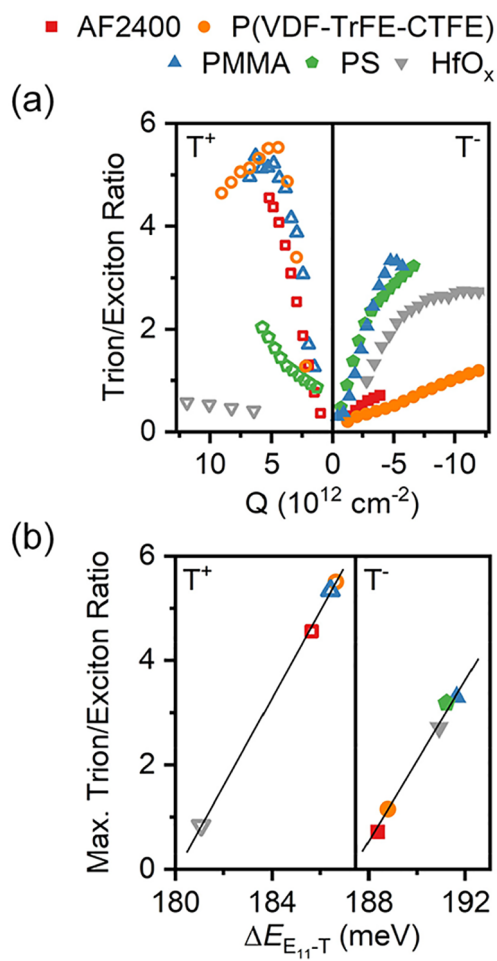

**Figure 6.** (a) Evolution of trion/exciton intensity ratios with charge carrier density $Q$. (b) Maximum trion/exciton intensity ratios vs corresponding exciton-trion energy separations with linear fits for hole (open symbols) and electron (solid symbols) accumulation.

An explanation for these unexpected trends might be given by the model introduced by Eckstein et al.[33,43] They proposed a hard segmentation of nanotubes by charge puddles to explain the emission properties of (electro)chemically doped (6,5) nanotubes. The assumption was that poorly screened counterions localize charge carriers and thus divide the nanotube into charge puddles, where trion emission originates from, and largely neutral nanotube segments where excitons are formed and reside. Diffusion of excitons along these segments and encounters with charge puddles lead to nonradiative decay. Consequently, higher doping levels and thus shorter neutral segments result in stronger exciton quenching. Exciton confinement in shorter segments also increases their energy, which should add to the expected blue-shift of the excitonic emission.

For a uniformly electrostatically doped system, most charge carriers should be highly mobile and not localized at particular positions because there are no adsorbed counterions as for electrochemically or chemically doped nanotubes. Hence, the idea of hard segmentation may not be applicable, and excitons should encounter mobile charges everywhere on the nanotubes with a high probability. Such a scenario should lead to strong exciton quenching, large trion/exciton ratios, and a pronounced blue-shift of the excitons as a result of shorter average diffusion distances, i.e., stronger axial confinement. However, when a given dielectric environment also leads to preferred trapping of one charge carrier, e.g., electrons for P(VDF-TrFE-CTFE) or holes for HfO$_x$, the creation of distinct charge puddles and hard segmentation of the nanotubes as proposed by Eckstein et al.[33] is again feasible. Compared to the case of mostly mobile charges, excitons would be less likely to interact with holes or electrons for the same charge carrier densities. This situation should lead to less exciton quenching, lower trion/exciton ratios, and a less pronounced exciton blue-shift. All of these predictions are qualitatively observed for the investigated FETs featuring gate dielectrics with different charge trapping properties. Overall, the localization of charges in trap states appears to reduce exciton quenching and trion formation.

Another contributing factor for the formation and overall concentration of trions might be their energetic stabilization over excitons as represented by the exciton-trion energy separation ($\Delta E_{E_{11} - T}$), which is always larger for negative trions (188−192 meV) compared to positive trions (181−187 meV), probably due to the slightly different effective masses of holes and electrons in small-diameter SWCNTs. No clear trend of $\Delta E_{E_{11} - T}$ with static dielectric constant is observed; however, a linear correlation of the maximum trion/exciton ratio with $\Delta E_{E_{11} - T}$ can be found for both positive and negative trions (see Figure 6b) with similar slopes of ∼0.8 per meV of $\Delta E_{E_{11} - T}$. The similar slopes indicate that these shifts should have the same physical origin, which may again point toward the effects of average/effective segmentation of the nanotubes by localized charges as proposed in the Eckstein model.[33] Further, $\Delta E_{E_{11} - T}$ at the maximum trion/exciton ratio increases roughly with the $E_{11}$ emission energy (see Supporting Information, Figure S9). However, a clear linear dependence as observed by Tanaka et al. resulting from the power law scaling of the interaction energies of both excitons and trions with dielectric constant[38] was not observed, and hence the physical origin of the scaling behavior remains unclear.

The experiments discussed above indicate that charge trapping is indeed detrimental to trion formation and reduces exciton quenching for the same overall charge carrier densities (trapped and mobile). The nonideal charge transport characteristics of the FETs, which vary with the different dielectrics, prevent a reliable quantitative analysis and comparison of the trap states as, for example, calculated from the subthreshold swings (see Methods). Nevertheless, in order to better understand the influence of trap states on the electrical and optical properties of SWCNTs, we used devices with a PMMA gate dielectric and reduced the density of electron traps by treatment of the nanotube networks with 1,2,4,5-tetrakis(tetramethylguanidino)benzene (ttmgb).[49,67] The guanidino-functionalized aromatic compound ttmgb is a two-electron donor (see Figure 7a). A layer of ttmgb decreases the work function of gold electrodes by about 1 eV, thus improving electron injection and blocking hole injection into (6,5) SWCNTs. As a strong base and reducing agent it also removes shallow electron traps caused by residual water and oxygen. To achieve lower trap densities, substrates with (6,5) SWCNT networks and patterned gold source-drain electrodes were dipped into ttmgb solutions in toluene with concentrations ($c_{ttmgb}$) of 0 (reference), 3, and 6 g·L$^{-1}$ prior to completion of the FETs with PMMA as the primary gate dielectric.

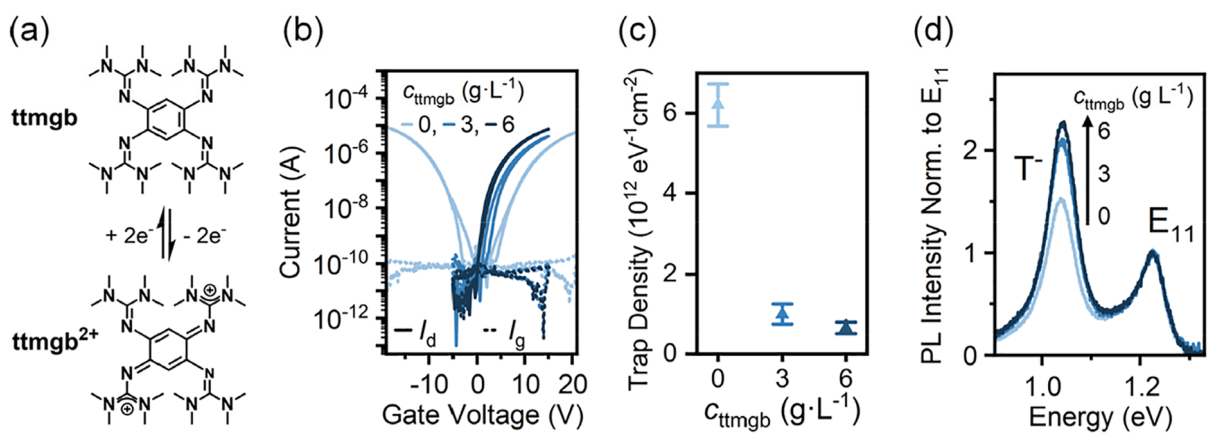

**Figure 7.** (a) Molecular structure of ttmgb and reversible two-electron oxidation/reduction. (b) Transfer characteristics ($L = 20$ μm, $W = 10$ mm, $V_d = 0.01$ V) of (6,5) SWCNT network FETs with PMMA dielectric. (c) Trap densities calculated from subthreshold swing, and (d) normalized PL spectra at the maximum trion/exciton ratios for different concentrations of ttmgb ($c_{ttmgb}$).

As expected, the devices treated with ttmgb exhibited purely n-type behavior without any hole transport, and the onset voltages shifted slightly with increasing ttmgb concentration (see Figure 7b). Analysis of the subthreshold swing of the transfer characteristics of the FETs (see Methods) revealed a substantial decrease of the trap densities in the ttmgb-treated devices as compared to the reference devices (see Figure 7c). Upon electron accumulation, the untreated (6,5) SWCNT networks showed the expected PL quenching, $E_{11}$ and trion blue-shift, and a peak in the trion/exciton ratio as discussed above. Note that the maximum trion/exciton ratio of the untreated PMMA sample was lower than for the PMMA devices in the dielectric comparison series. However, the corresponding $\Delta E_{E_{11} - T}$ value fits well within the linear trend for negative trions in Figure 6b (see Supporting Information, Figure S10a), indicating that this correlation is universal. Upon ttmgb treatment, exciton quenching, blue-shift, and trion/exciton ratio evolution with carrier concentration were observed as before. The absolute maximum trion/exciton ratios increased compared to the pristine networks as expected for fewer localized electron traps (see discussion above). However, these maxima were also reached at higher electron densities probably due to the larger share of mobile carriers (see Figure 7d and Supporting Information, Figure S10b). The observed effects of ttmgb treatment of the nanotube network are moderate but consistent with the proposed model. The data also implies that the suspected charge localization effects of the dielectrics are stronger than those of network- or substrate-intrinsic charge traps.

## CONCLUSIONS

In summary, we systematically investigated the impact of the dielectric environment and associated charge traps on trion and exciton emission in electrostatically doped (6,5) single-walled carbon nanotube networks in field-effect transistors. The exciton energies in the neutral state roughly scaled with the polarity and dielectric constant of the gate dielectrics. However, exciton quenching efficiency, exciton blue-shifts, maximum trion/exciton intensity ratios, and exciton-trion energy separations upon electrostatic doping varied strongly with the dielectric and specifically with the carrier polarity for the same dielectric. These variations could not be explained by the different polarizabilities and charge screening characteristics of the gate dielectrics, but they correlated well with their charge trapping properties. Gate dielectrics that showed predominant electron trapping (i.e., fluorinated polymers) resulted in significantly less exciton quenching and $E_{11}$ blue-shift as well as lower maximum trion to exciton intensity ratios for electron accumulation compared to hole accumulation. For HfO$_x$ as a gate dielectric with hole trapping properties these effects were reversed. The localization (trapping) of charge carries, and thus the formation of charge puddles with trion emission separated from neutral nanotube segments with exciton emission[33] as opposed to the interaction (quenching and trion formation) of fully mobile charge carriers with highly diffusive excitons may explain this observation. Consequently, the controlled introduction of charge traps might be a possible route for the reduction of unwanted trion emission. However, the associated reduction of carrier mobility may outweigh the possible gains for optoelectronic devices. Overall, we conclude that exciton and trion emissions from doped carbon nanotubes are highly dependent on the precise dielectric environment and carrier type. At least in complex device structures such as FETs no simple correlation with static dielectric constants can be made, but charge traps must be considered as well. Furthermore, neither the ratio of trion to exciton emission nor the observed exciton blue-shift can be used as general quantitative metrics for the carrier density.

## ■ ASSOCIATED CONTENT

**Ⓢ Supporting Information**

The Supporting Information is available free of charge at https://pubs.acs.org/doi/10.1021/acs.jpcc.2c08338.

> Raman and absorption spectra of (6,5) SWCNT dispersions, atomic force micrographs of the polymeric dielectric layers, additional electrical characterization (areal capacitances, output curves, current hysteresis), local PL inhomogeneities of nanotube networks and removal by large area measurements, details of PL quenching analysis (onset voltage shifts, spectral fitting, $E_{11}$ peak position and line width analysis, charge carrier density-dependent emission energies and trion/exciton ratios, maximum $E_{11}$ peak shifts, trion-exciton energy separation vs $E_{11}$ peak position), further analysis of trion/exciton ratios upon ttmgb treatment (PDF)

■ AUTHOR INFORMATION

**Corresponding Author**

Jana Zaumseil − *Institute for Physical Chemistry, Universität Heidelberg, D-69120 Heidelberg, Germany*; orcid.org/0000-0002-2048-217X; Email: zaumseil@uni-heidelberg.de

**Authors**

Sonja Wieland − *Institute for Physical Chemistry, Universität Heidelberg, D-69120 Heidelberg, Germany*

Abdurrahman Ali El Yumin − *Institute for Physical Chemistry, Universität Heidelberg, D-69120 Heidelberg, Germany*

Jan M. Gotthardt − *Institute for Physical Chemistry, Universität Heidelberg, D-69120 Heidelberg, Germany*

Complete contact information is available at:
https://pubs.acs.org/10.1021/acs.jpcc.2c08338

**Notes**

The authors declare no competing financial interest.

■ ACKNOWLEDGMENTS

This project has received funding from the European Research Council (ERC) under the European Union's Horizon 2020 research and innovation programme (Grant No. 817494 "TRIFECTs"). The authors thank H.-J. Himmel and his group for provision of ttmgb.

# Supporting Information

Impact of Dielectric Environment on Trion Emission from Single-Walled Carbon Nanotube Networks

*Sonja Wieland, Abdurrahman Ali El Yumin, Jan M. Gotthardt, Jana Zaumseil\**

Institute for Physical Chemistry, Universität Heidelberg, D-69120 Heidelberg, Germany

**Corresponding Author**

*\*zaumseil@uni-heidelberg.de*

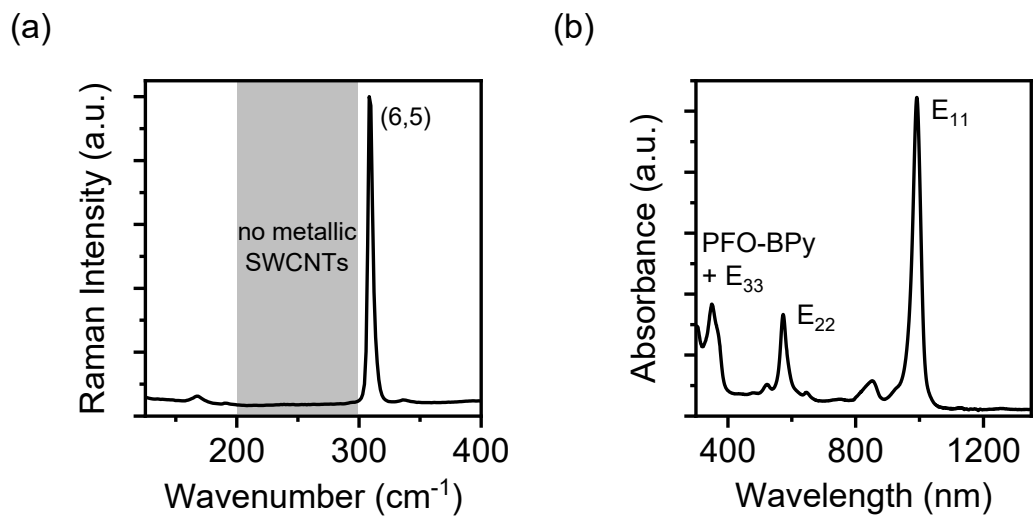

**Figure S1.** (a) Raman spectrum of dropcast (6,5) SWCNTs from stock dispersion (radial breathing mode range, excitation wavelength 532 nm) and (b) UV-Vis-nIR absorption spectrum of nanotube dispersion in toluene after polymer-sorting and removal of excess PFO-BPy. Both corroborate the highly selective dispersion of (6,5) SWCNTs and absence of metallic or other nanotube species.

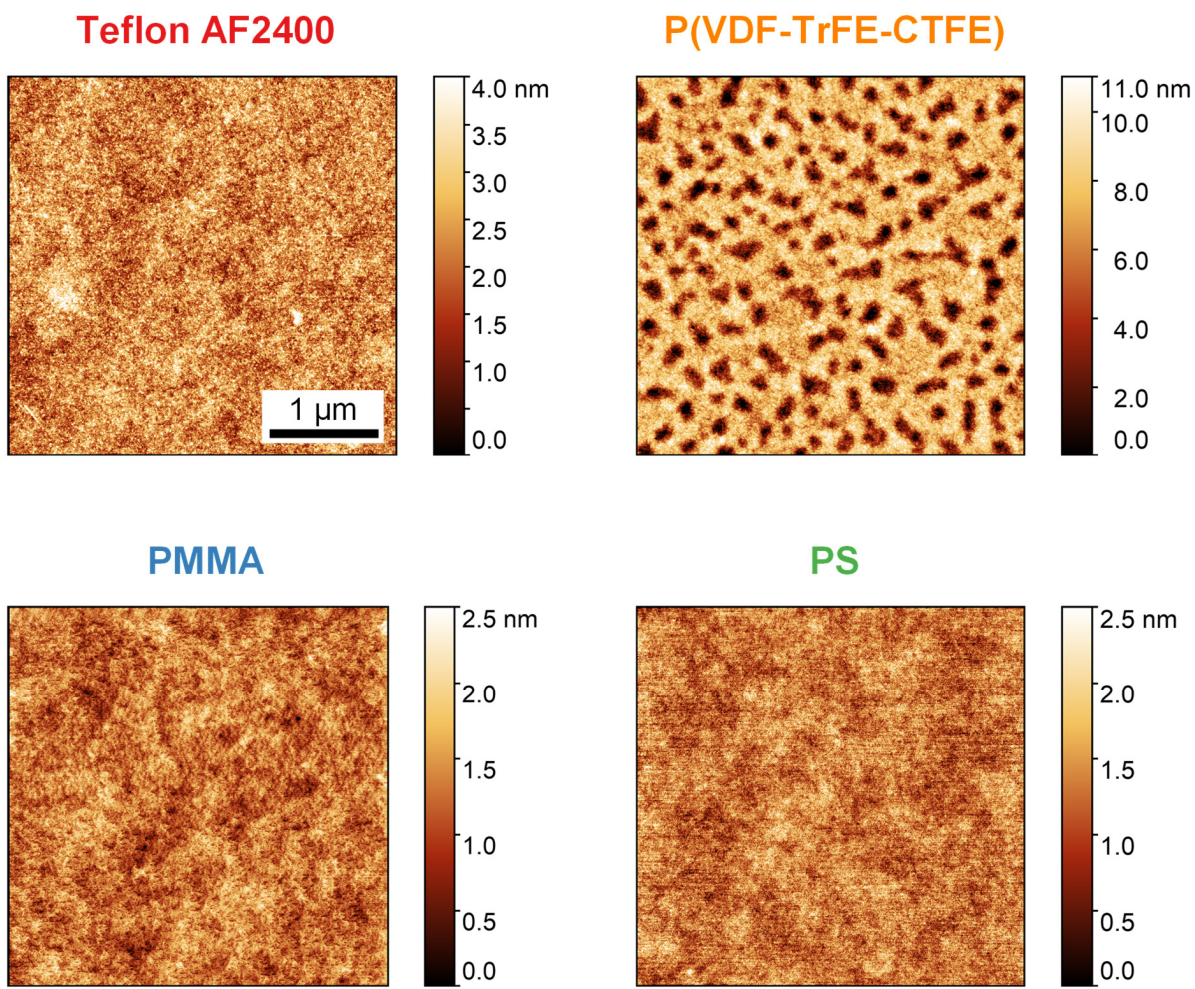

**Figure S2.** Atomic force microscopy (AFM) images of spincoated polymer layers on glass. The polymers formed homogeneous and smooth films (root mean square roughness 0.67 nm, 0.38 nm and 0.41 nm for AF2400, PMMA and PS, respectively) except copolymer P(VDF-TrFE-CTFE), for which some phase segregation was observed (root mean square roughness 2.3 nm).

**Table S1.** Areal capacitances for each (hybrid) gate dielectric as measured on FETs at 1000 Hz.

| | AF2400 | P(VDF-TrFE-CTFE) | PMMA | PS | HfO$_x$ |
|---|---|---|---|---|---|
| Areal capacitance (nF·cm$^{-2}$) | 26 | 123 | 39 | 32 | 198 |

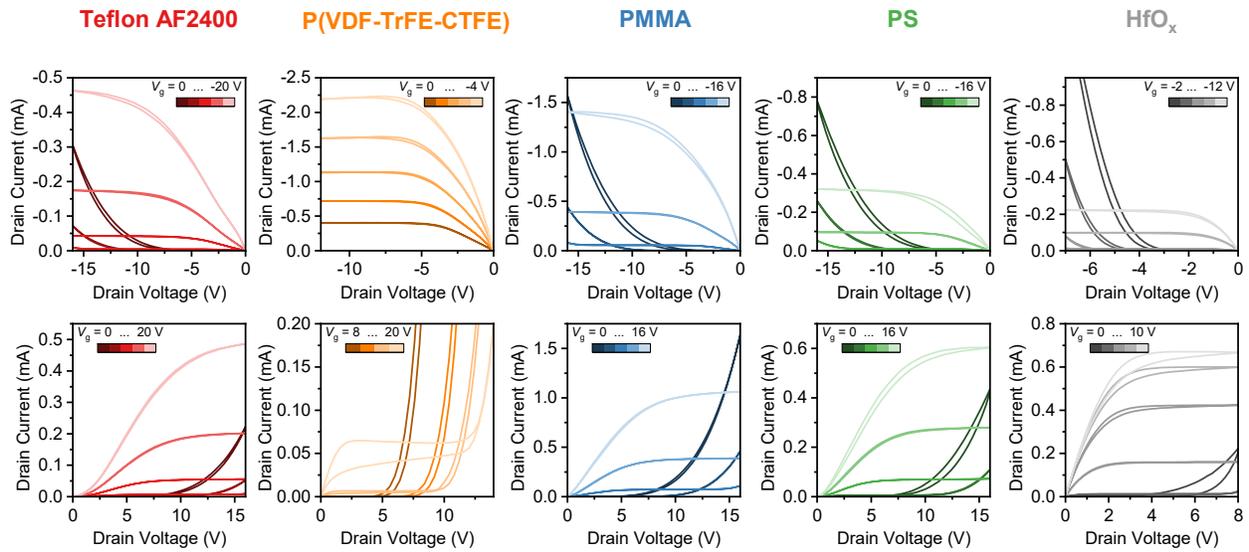

**Figure S3.** Output curves of FETs with different (hybrid) dielectrics (channel length $L$ = 40 μm, channel width $W$ = 5 mm) for hole injection (top row) and electron injection (bottom row). Non-Ohmic contacts observed for some devices suggest that the gate dielectric also influences charge injection.

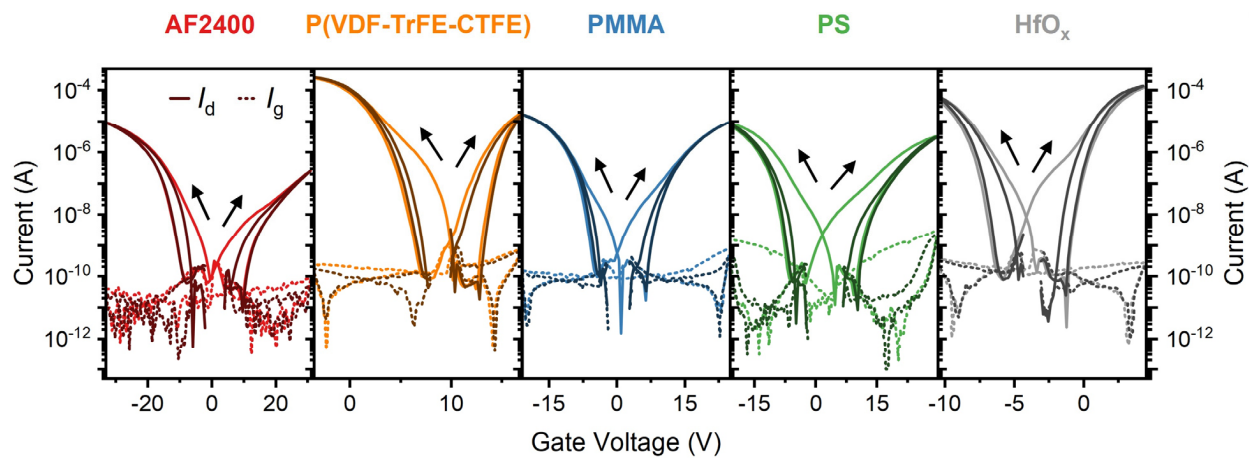

**Figure S4.** Dependence of current hysteresis in transfer characteristics on gate voltage range at low drain voltages ($V_d$ = -0.01 V) for FETs with different gate dielectrics in contact with the SWCNTs ($L$ = 10 μm, $W$ = 10 mm). The gate voltage range was adjusted to induce both hole and electron accumulation within one $V_g$ sweep (full range, bright lines), or only one carrier type (limited range, dark lines). Dashed lines indicate corresponding gate leakage currents.

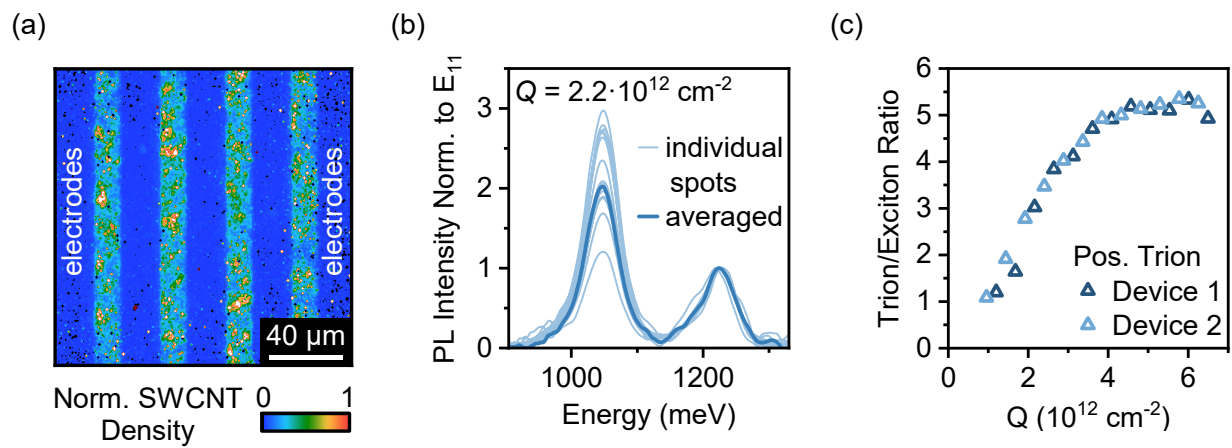

**Figure S5.** Dependence of the trion/exciton ratios on local (6,5) SWCNT density, as demonstrated for an FET with PMMA as the gate dielectric and for hole accumulation. (a) Image of photoluminescence (PL) intensity (excitation wavelength 640 nm) at minimum charge carrier density $Q$ within the transistor channel (section of an area including interdigitated source-drain electrodes) showing non-uniform spatial distribution of (6,5) SWCNTs. (b) Smoothed PL spectra of individual spots and corresponding averaged spectrum. The observed trion/exciton ratio at a certain carrier density for the entire device (based on gate voltage) depends on the local SWCNT density, leading to spot-to-spot variations. (c) Reproducible dependence of positive trion/exciton ratio on overall carrier density for different devices obtained after averaging over an area of 700 μm².

**Table S2.** Gate voltages applied to reach maximum PL intensity, *i.e.*, the least charged state (defined here as $Q = 0$).

|  | AF2400 | P(VDF-TrFE-CTFE) | PMMA | PS | HfO$_x$ |
|---|---|---|---|---|---|
| $V_g(Q=0)$ (V) | 2 | 7.5 | 0 | -2 | -2 |

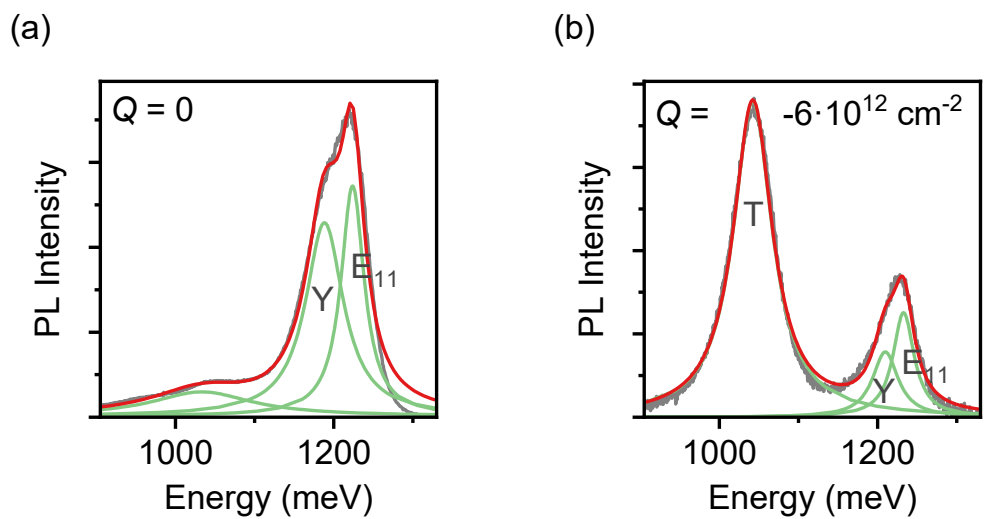

**Figure S6.** Multi-peak Lorentz fit of PL spectra of (6,5) SWCNT network in an FET with a PMMA gate dielectric (a) at zero and (b) at high charge carrier density (electrons). In the uncharged state, the sideband in the low-energy range of the spectrum could not be neglected in the fitting procedure.

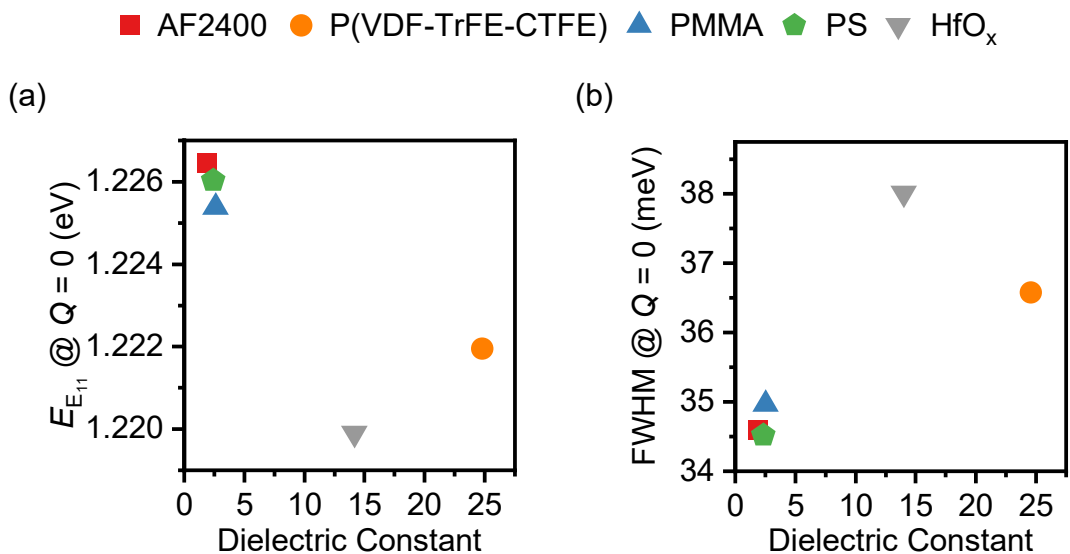

**Figure S7.** (a) Fitted $E_{11}$ emission peak energy and (b) full width at half maximum (FWHM) of $E_{11}$ emission peak in the uncharged state depending on static dielectric constant of gate dielectric.

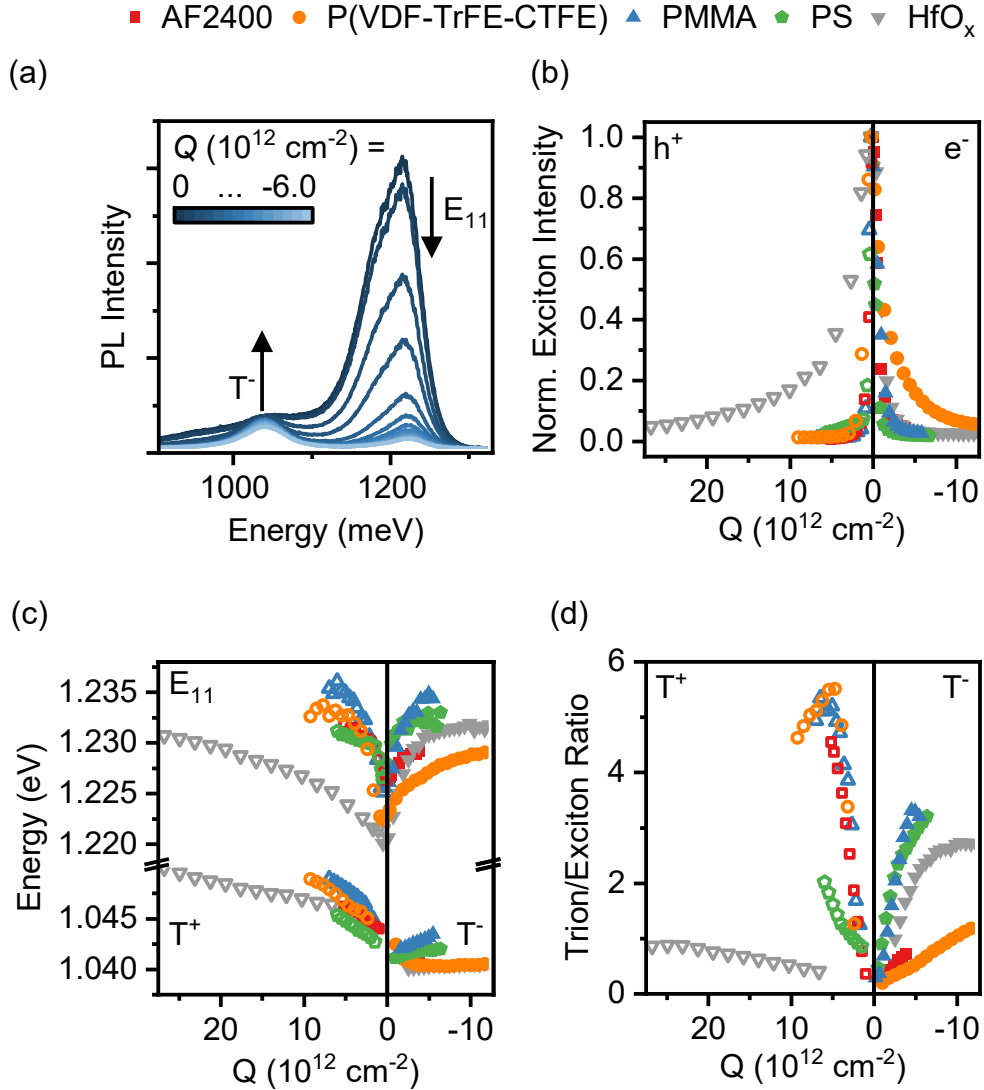

**Figure S8.** Impact of charge carrier density on excitons and trions. (a) Evolution of (6,5) SWCNT network PL spectrum with carrier density ($Q$ in charges per cm²) for electron accumulation (PMMA gate dielectric). (b) Exciton emission intensity depending on $Q$ for different gate dielectrics and for hole (open symbols) and electron (solid symbols) accumulation. Complete $Q$ range of charge carrier density dependence of (c) $E_{11}$ and trion emission peak energies and (d) trion/exciton intensity ratios.

**Table S3.** Maximum $E_{11}$ peak blue-shifts (in meV) upon hole and electron accumulation.

|  | **AF2400** | **P(VDF-TrFE-CTFE)** | **PMMA** | **PS** | **HfO$_x$** |
|---|---|---|---|---|---|
| hole accumulation | 5.5 | 11.1 | 10.1 | 4.6 | 9.3 |
| electron accumulation | 2.7 | 6.7 | 9.4 | 6.9 | 11.9 |

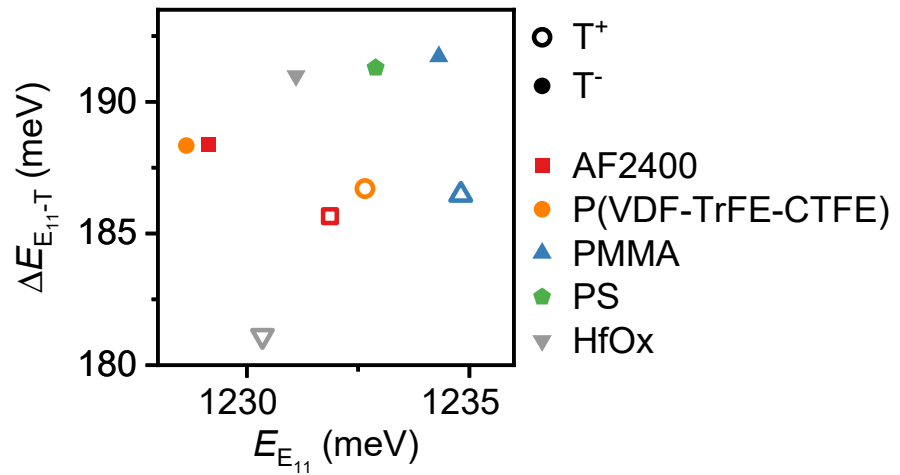

**Figure S9.** Energy separation between excitons and trions ($\Delta E_{E_{11}-T}$) at the maximum trion/exciton ratio *versus* $E_{11}$ emission peak energy for positive (open symbols) and negative (solid symbols) trions.

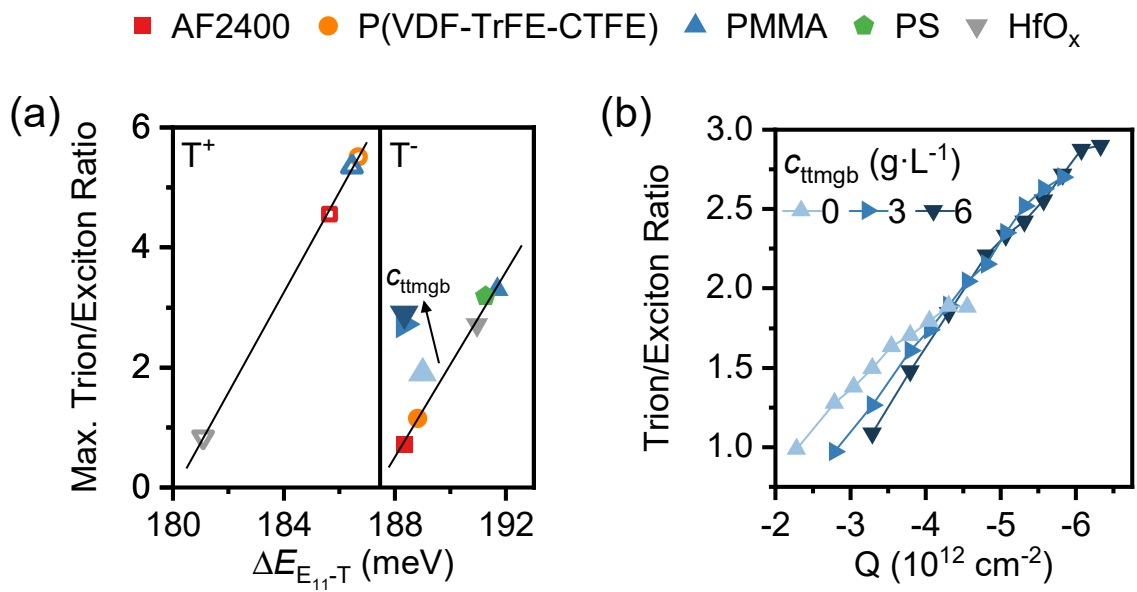

**Figure S10.** (a) Maximum trion/exciton ratios at the corresponding exciton-trion energy separation ($\Delta E_{E_{11}-T}$) with linear fits for different dielectrics in hole (open symbols) and electron (solid symbols) accumulation. The results of the ttmgb-doping experiments for devices with PMMA dielectric are added as large blue triangles. (b) $Q$-dependent trion/exciton ratios in ttmgb-doped devices (lines are guides for the eye).